\documentclass[twocolumn,nolinenumbers,trackchanges]{aastex631}

\shorttitle{Exploring the hadronic origin of LHAASO J1908+0621}
\shortauthors{De Sarkar $\&$ Gupta}
\graphicspath{{./}{figures/}}
\usepackage{amsmath}
\usepackage{hyperref}
\usepackage{xcolor}
\begin{document}

\title{Exploring the hadronic origin of LHAASO J1908+0621}

\author[0000-0001-6047-6746]{Agnibha De Sarkar}
\affiliation{Astronomy $\&$ Astrophysics group, Raman Research Institute \\
C. V. Raman Avenue, 5th Cross Road, Sadashivanagar, Bengaluru 560080, Karnataka, India}
\email{agnibha@rri.res.in}

\author[0000-0002-1188-7503]{Nayantara Gupta}
\affiliation{Astronomy $\&$ Astrophysics group, Raman Research Institute \\
C. V. Raman Avenue, 5th Cross Road, Sadashivanagar, Bengaluru 560080, Karnataka, India}

\begin{abstract}

 Recent observations by the Large High Altitude Air Shower Observatory (LHAASO) have paved the way for the observational detection of PeVatrons in the Milky Way Galaxy, thus revolutionizing the field of $\gamma$-ray astrophysics. In this paper, we study one such detected source, LHAASO J1908+0621, and explore the origin of multi-TeV $\gamma$-ray emission from this source. A middle-aged radio supernova remnant SNR G40.5-0.5 and a GeV pulsar PSR J1907+0602 are co-spatial with LHAASO J1908+0621. Dense molecular clouds are also found to be associated with SNR G40.5-0.5. We explain the multi-TeV $\gamma$-ray emission observed from the direction of LHAASO J1908+0621, by the hadronic interaction between accelerated protons that escaped from the SNR shock front and cold protons present inside the dense molecular clouds, and the leptonic emission from the pulsar wind nebula (PWN) associated with the pulsar J1907+0602. Moreover, we explain lower energy $\gamma$-ray emission by considering the radiative cooling of the electrons that escaped from SNR G40.5-0.5. Finally, the combined lepto-hadronic scenario was used to explain the multi-wavelength spectral energy distribution (SED) of LHAASO J1908+0621. Although not yet significant, an ICECUBE hotspot of neutrino emission is spatially associated with LHAASO J1908+0621, indicating a possible hadronic contribution. In this paper, we show that if a hadronic component is present in LHAASO J1908+0621, then the second generation ICECUBE observatory will detect neutrino from this source.
\end{abstract}


\section{Introduction} \label{sec:intro}

Cosmic rays (CR) are charged atomic nuclei, traversing through the space with relativistic speed. The CRs consist of 90 $\%$ protons, about 8-9 $\%$ Helium nuclei and smaller abundances of heavier elements. The observed local proton spectrum can be well described by a single power law with an index of -2.7, up to around 1 PeV (= 10$^{15}$ eV) energy, which is also known as the ``knee'' of the CR spectrum. This hints towards the presence of powerful astrophysical proton accelerators in our Galaxy, which can accelerate the CR protons to PeV energies, the so-called ``PeVatrons''. Despite having been  theoretically studied very thoroughly, no Galactic source has been unambiguously confirmed to be a PeVatron, except the possible case of the Galactic center \citep{hess18, hess20, magic20}. Since CRs, which can accelerate up to PeV energies, can interact with ambient medium to produce multi-TeV energy $\gamma$-rays, the PeVatrons can be identified by studying the association of $\gamma$-ray sources with them. To that end, successful operations by ground based observatories such as \href{https://www.mpi-hd.mpg.de/hfm/HESS/}{H.E.S.S.} (High Energy Stereoscopic System), \href{https://magic.mpp.mpg.de/}{MAGIC} (Major Atmospheric Gamma Imaging Cherenkov), \href{https://www.icrr.u-tokyo.ac.jp/em/index.html}{Tibet AS$\gamma$}, \href{https://www.hawc-observatory.org/}{HAWC} (High-Altitude Water Cherenkov), \href{http://english.ihep.cas.cn/lhaaso/}{LHAASO} (Large High Altitude Air Shower Observatory) over the past ten years have made the ultra-high energy (UHE) $\gamma$-ray astronomy an active area of research. Since the ultra-high energy $\gamma$-rays produced outside our Galaxy get heavily attenuated by the Cosmic Microwave Background (CMB) and Infrared Background (IRB), it is difficult to detect UHE $\gamma$-ray sources outside our Galaxy. However, the recent detections of many $\gamma$-ray sources, emitting $\gamma$-rays with energies ranging from several hundreds of TeV to PeV, have increased the possibilities to unambiguously confirm the presence of PeVatrons residing in our Galaxy.

LHAASO is a state-of-the-art dual-task facility designed for CRs and $\gamma$-ray studies at few hundred GeV to few PeV, located at 4410 m above sea level in China \citep{lhaaso10}. Since starting its operation from April 2020, LHAASO has detected more than a dozen of UHE $\gamma$-ray sources in our Galaxy. Many of these sources are associated with PWNe or SNRs. Since LHAASO obervatory is sensitive enough to detect UHE $\gamma$-rays coming from a source, the chances of establishing astrophysical sources such as PWN or SNR as possible PeVatrons are very strong. For this work, we study one of such UHE $\gamma$-ray sources observed by LHAASO, which has a strong possibility of being a Galactic PeVatron \citep{Cao21}.

LHAASO J1908+0621 is a UHE $\gamma$-ray source, detected in a serendipitous search for $\gamma$-ray sources by LHAASO observatory \citep{Cao21}. This source was detected with 12 other sources with energies $\geq$ 100 TeV and statistical significance $\geq$ 7$\sigma$. The LHAASO source is located at RA = 287.05$^{\circ}$ and Decl. = 6.35$^{\circ}$, with a significance above 100 TeV to be 17.2$\sigma$, making it one of the brightest UHE $\gamma$-ray source in our Galaxy. The $\gamma$-ray spectrum of this source reaches up to a maximum energy of 0.44 $\pm$ 0.05 PeV, and the differential photon flux of this source at 100 TeV was found to be 1.36 $\pm$ 0.18 Crab Unit (Crab Unit = flux of the Crab nebula at 100 TeV, 1 Crab Unit = 6.1 $\times$ 10$^{-17}$ photons TeV$^{-1}$ cm$^{-2}$ s$^{-1}$). Since the maximum energy the UHE $\gamma$-rays emitted from LHAASO J1908+0621 can attain is greater than 100 TeV, this source shows a strong possibility of being associated with a Galactic PeVatron.

\cite{Cao21} has obtained and fitted the $\gamma$-ray spectrum of LHAASO J1908+0621 with a simple power law model and a log parabola model. The log parabola model gives a better fit compared to a simple power law model, due to a gradual steepening of $\gamma$-ray spectrum between 10 TeV and 500 TeV. Although this steepening can be due to $\gamma$-ray absorption from background photons, the effect of absorption was found to be small, even at very high energies. The best-fit parameters for the log-parabola $\gamma$-ray spectral fit of LHAASO J1908+0621 are a = 2.27 and b = 0.46, where log parabola model is defined by (E/10 TeV)$^{-a-b\:log(E/10 TeV)}$. The 68$\%$ contamination angle for LHAASO J1908+0621 was found to be 0.45$^{\circ}$, obtained for $\gamma$-rays over 25 TeV. HAWC observatory has observed the LHAASO source to have a hard spectrum reaching energies above 100 TeV without any hint of an exponential cutoff, making it a best case for Galactic PeVatrons \citep{hawc19}. This source was first observed by MILAGRO observatory \citep{abdo07}, and was later confirmed by H.E.S.S. observatory \citep{aharo09}, which detected the source with large angular size ($\sigma$ = 0.34$^{\circ}$) and a hard spectral index of 2.1, above 300 GeV. ARGO-YBJ observatory  \citep[Astrophysical Radiation with Ground-based Observatory at YangBaJing;][]{bartoli12} has found that the TeV luminosity of this source is comparable to Crab Nebula, which makes it one of the most luminous Galactic $\gamma$-ray sources in the TeV regime. The observation by VERITAS \citep[Very Energetic Radiation Imaging Telescope Array System;][]{aliu14} observatory has also revealed an extended source system ($\sigma$ = 0.44$^{\circ}$), with three peaks of emission and also, a photon index of 2.2. Additionally, the LHAASO source is associated with an ICECUBE neutrino hotspot, although the significance is low \citep{icecube19, icecube20}. The extended nature of the LHAASO source indicates that a SNR and/or PWN should be associated with this source. To that end, the study of possible counterparts of LHAASO J1908+0621 is necessary to establish both $\gamma$-ray production region and nearby particle accelerators.

LHAASO J1908+0621 is spatially associated with a middle aged, shell-type supernova remnant SNR G40.5-0.5 \citep[20-40 kyr;][]{downes}, which is brighter in the northern region of the TeV source, as observed in the radio data obtained by VLA Galactic Plane Survey \citep[VGPS;][]{stil}. The distance estimation places the SNR at a distance of 3.5 kpc, by CO observations \citep{yang06} or a more distant position of 5.5-8.5 kpc, using $\Sigma$-D relation \citep{downes} and 6.1 kpc \citep{case}. The recently discovered relatively young and energetic radio pulsar PSR J1907+0631 (characteristic age $\tau$ = 11 kyr, spin-down luminosity $\sim$ 5 $\times$ 10$^{35}$ erg s$^{-1}$) lies close to the projected center of the SNR \citep{lyne}. The estimated distance of this pulsar obtained from dispersion measure (DM) is 7.9 kpc, which is compatible with the estimated distance of G40.5-0.5 and hints towards an association between these two objects. Although, in principle, PSR J1907+0631 can power the entire TeV source \citep{dudi20}, the considerable offset between the pulsar and the position of the $\gamma$-ray emission disfavours that scenario. Additionally, distribution of molecular clouds (MCs) has been confirmed from studies involving the distribution of CO gas in the vicinity of SNR G40.5-0.5. \cite{Li21} have searched for MCs with $^{12}$CO (J=1-0), $^{13}$CO (J=1-0) and C$^{18}$O (J=1-0) emission lines, and discovered the MCs to be spatially associated with SNR G40.5-0.5 in the $^{12}$CO (J=1-0) and $^{13}$CO (J=1-0) maps between the integrated velocity range of 46 and 66 km s$^{-1}$. A shell-like cavity around the radio morphology of SNR G40.5-0.5 was also observed in the $^{12}$CO (J=1-0) and $^{13}$CO (J=1-0) maps, indicating a possible SNR swept-up shell \citep{Li21}. The presence of MCs is also confirmed by \cite{crestan}, in which they were discovered in the $^{12}$CO (J=1-0) and $^{13}$CO (J=1-0) maps in the velocity range of 58-62 km s$^{-1}$. This discovery places the SNR+MC association at a near distance of $\sim$ 3-3.5 kpc and far distance of $\sim$ 8-9.5 kpc, and the corresponding mean number density of the MCs were estimated to be 110-180 cm$^{-3}$ assuming near distance and 45-60 cm$^{-3}$ assuming far distance \citep{Li21, crestan}.

Apart from the SNR G40.5-0.5 and PSR J1907+0631, a $\gamma$-ray loud pulsar, PSR J1907+0602, was also found to be spatially associated with LHAASO J1908+0621, located in the southern part of the source \citep{crestan}. The pulsar has a characteristic age of 19.5 kyr and a spin-down luminosity of 2.8 $\times$ 10$^{36}$ erg s$^{-1}$ \citep{abdo10}. The distance of the pulsar, estimated from DM, was found to be 3.2 $\pm$ 0.6 kpc \citep{abdo10}. \cite{Li21} performed  off-pulse analysis of the \textit{Fermi}-LAT data of the GeV pulsar PSR J1907+0602, and discovered a previously undetected, extended source spatially associated with the Milagro counterpart of LHAASO J1908+0621, labeled as Fermi J1906+0626. Additionally, another unidentified GeV source 4FGL J1906.2+0631, distance unknown, was located within the positional error of LHAASO J1908+0621 \citep{abdollahi20}. 

Due to its complex spatial morphology, the origin of the $\gamma$-ray emission from LHAASO J1908+0621 is uncertain. Leptonic emission from PWN associated with PSR J1907+0602 can be a possible origin of the multi-TeV $\gamma$-ray detected by LHAASO. The electrons can be accelerated  up to 1 PeV at the wind termination shock of the PWN. However, electrons being leptons, lose energy radiatively very fast. Thus escape from the acceleration site and then further propagation can pose a real challenge to the scenario \citep{Cao21}. Furthermore, if electrons are the progenitor of the $\gamma$-ray emission, no neutrinos should be detected by ICECUBE from the source region and the neutrino hotspot could not be explained. Alternatively, escaped protons from shock of the SNR G40.5-0.5 can penetrate the associated MCs and through hadronic interaction, produce multi-TeV $\gamma$-rays. Although the SNR itself is too old to produce multi-TeV $\gamma$-rays, protons accelerated at earlier epochs can initiate high energy $\gamma$-ray emission from the MC region \citep{Cao21}. Moreover, in the hadronic scenario, one can also explain the neutrino hotspot near the source region. Intrigued by this fact, in this work, we explore the hadronic origin of LHAASO J1908+0621. We try to see the conditions in which the emission from the LHAASO source can be explained by $\gamma$-rays originated from p-p interaction between accelerated protons from the SNR and cold protons inside the MCs, as well as calculate the corresponding neutrino emission from the hadronic interaction and compare it to the sensitivity of the ICECUBE Gen-2 observatory \citep{icecubegen2}. Additionally, we also consider leptonic emission from the PWN associated with PSR J1907+0602, as well as the leptonic emission from the SNR+MC system, along with the hadronic contribution, to understand the radiation mechanism implied by the observed multi-wavelength (MWL) SED.

In section \ref{sec:mor}, we discuss the morphology of the complicated region surrounding LHAASO J1908+0621. In section \ref{sec:had}, we calculate the multi-TeV $\gamma$-ray emission through hadronic interaction between accelerated protons from SNR and cold protons residing in the MCs. In section \ref{sec:lep}, we calculate the leptonic contributions from both SNR G40.5-0.5 and PWN associated with PSR J1907+0602. In section \ref{sec:neu}, we calculate the corresponding neutrino SED from the hadronic interaction, and compare the calculation with the sensitivity of ICECUBE observatory. In section \ref{sec:DC}, we discuss the obtained results, and in section \ref{sec:CON}, we conclude this work.  

\section{Morphology} \label{sec:mor}

Detailed morphological study of the region surrounding LHAASO J1908+0621 has been reported in \cite{Li21}, \cite{crestan} and \cite{Cao21}, using various observations by space-based and ground-based observatories. Through detailed \textit{Fermi}-LAT data analysis, \cite{Li21} has reported the position of the PWN associated with PSR J1907+0602. The position of SNR G40.5-0.5 and the surrounding MCs were also confirmed by radio observations and CO mapping respectively \citep{Li21, crestan}. In general, a clear separation of high energy radiation from the low energy emission, attributed to their different original objects, must be strongly supported by the morphological observation of the extended source, in which the objects are spatially well separated. However, due to the complex juxtaposition of potential counterparts along the line-of-sight of LHAASO J1908+0621, it is difficult to distinguish among the sources responsible for high energy and low energy emissions from the region. In this section, we discuss the emission mechanisms considered in this paper, to explain the MWL SED of LHAASO J1908+0621, while being consistent with the observed energy morphology of the source region.

Ground-based observatories, such as H.E.S.S. and VERITAS have good enough angular resolutions ($\sim$ 0.06$^\circ$) to extract a high energy emission region in the direction of the LHAASO source. Although the significance is not very high, the significance map derived from the VERITAS observation, indicates that the PWN associated with PSR J1907+0602 could be an important source for the VHE (E $>$ 100 GeV) $\gamma$-ray emission \citep{aliu14, Li21,crestan}. However, as given in \cite{crestan}, although the VERITAS emission lobe obtained from the significance map (significance levels ranging from 3$\sigma$ to 5.2$\sigma$), matches well with the proposed PWN position, there is another VERITAS emission lobe, which is spatially coincident with the contact point between SNR G40.5-0.5 and the surrounding MCs (see Figure 1 and 3 of \cite{crestan}). This indicates that VHE $\gamma$-rays obtained from both PWN J1907+0602 and the SNR+MC system should contribute to the SED obtained by VERITAS. Similar to VERITAS, in the H.E.S.S. significance map obtained by H.E.S.S. Galactic Plane Survey \citep[HGPS;][]{hgps18}, emission lobes were found to be coincident with the position of PWN J1907+0602, as well as the contact region of the SNR+MC system. Moreover, the 68$\%$ containment region of the Gaussian morphology measured by H.E.S.S., comfortably overlaps with the SNR+MC system, as well as the PWN. The fact that the H.E.S.S. energy morphology contains two emission lobes spatially attributed to the VHE $\gamma$-ray radiation from  both the PWN and the SNR+MC system, suggests that emissions from both the PWN and the SNR+MC system should be responsible for the VHE $\gamma$-ray data observed by H.E.S.S \citep{aharo09,hgps18}. This is why we have considered the contributions from both PWN J1907+0602 and the SNR+MC system to satisfy the VHE SEDs obtained from VERITAS and H.E.S.S. observations. We note that since the Field of View (FoV) of VERITAS observatory \citep[FoV $\sim$ 3.5$^\circ$;][]{galante12} is smaller than that of H.E.S.S. observatory \citep[FoV $\sim$ 5$^\circ$;][]{DeNaurois}, VERITAS underestimates the total flux observed from the source region, as compared to that measured by H.E.S.S. observatory. We have taken this into account while constructing the MWL SED of the source, and we have scaled the VERITAS SED to match that measured by H.E.S.S. observatory.

For other ground-based observatories such as LHAASO and HAWC, the angular resolution may not be enough to draw a detailed morphological map of the source region, but they should be enough to establish the extent of high energy emission from the source region. \cite{Cao21} has provided KM2A significance map, which shows the potential counterparts of the UHE ( $>$ 100 TeV) $\gamma$-ray source. From the inset of the Extended data Figure 5 of \cite{Cao21}, it can be seen that, similar to H.E.S.S., the reported PWN position given by \cite{Li21} is within the extent of the UHE emission observed by LHAASO, although an offset of 0.18$^\circ$, or 10 pc (at 3.2 kpc), is also present between the centroid of the LHAASO emission morphology and the best-fit position of the disk morphology used to explain the PWN in \cite{Li21}. On the other hand, the overlapping region of SNR G40.5-0.5 and the surrounding MCs is also situated well within the maximum significance region observed by LHAASO, however the centroid of the UHE emission morphology observed by LHAASO, is also not coincident with the contact region between SNR G40.5-0.5 and the associated MCs. No distinct lobes of emission, like in the cases of H.E.S.S. and VERITAS, were found in the source morphology observed by LHAASO, making it difficult to ascertain which source, the PWN or the SNR+MC system, is actually contributing in the UHE regime. \cite{hawc21} has tried to explain the UHE $\gamma$-ray data observed by LHAASO using one-zone and two zone, purely leptonic scenarios originated from PWN J1907+0602. However, it was found that the corresponding synchrotron fluxes obtained from the proposed models exceed the X-ray upper limits measured by XMM-Newton observatory \citep{hawc21}. Consequently, in this work we explore the contribution of the SNR+MC system, which is another possible candidate overlapped with the image of LHAASO J1908+0621 \citep{Cao21}, and determine the conditions for which the SNR+MC system would be responsible for the UHE $\gamma$-ray emission observed by LHAASO. The phenomenological model explored in this paper does not violate the observed X-ray upper limits. 

As stated earlier, \cite{Li21} had discovered an extended source by performing \textit{Fermi}-LAT data analysis during the off-peak phases of the PSR J1907+0602. They have shown that this extended source, Fermi J1906+0626, shows a significant peak coincident with the molecular material distribution obtained from the CO mapping. This clearly implies that Fermi J1906+0626 is a result of interaction between accelerated particles from SNR G40.5-0.5 and the associated MCs. Moreover, as seen from the Figure 2 of \cite{Li21}, the significance peak of Fermi J1906+0626 in 0.1 - 2 GeV energy range, is outside of the TeV significance contours presented by VERITAS, as well as UHE emission morphology of LHAASO. So in this work, we explain the lower energy emission from the direction of LHAASO J1908+0621, by the leptonic interaction between escaped electrons from the SNR shock front and the molecular material. Since inside MCs, bremsstrahlung radiation will dominate IC cooling, due to enhanced material number density, we have explained the lower energy $\gamma$-ray SED (0.1 - 10 GeV) using bremsstrahlung emission in the present work. Below, we discuss the theoretical framework of our model, used to explain the MWL SED of LHAASO J1908+0621.

\section{Hadronic modeling} \label{sec:had}

\begin{figure*}[ht!]
\plotone{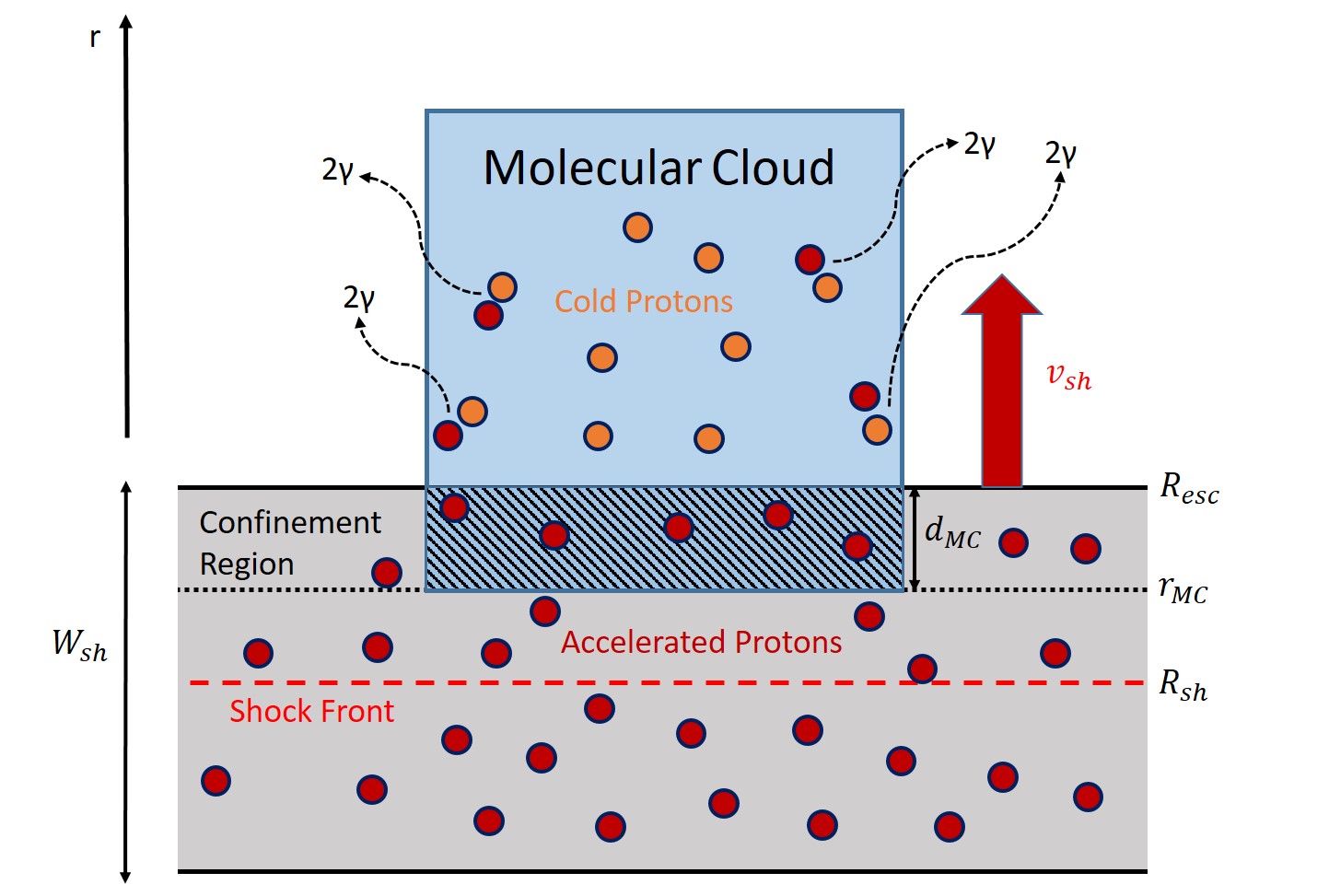}
\caption{\label{fig1} Schematic diagram showing interaction between the SNR and associated MCs, following \cite{makino}. The radially outward direction from the cavity center is signified by the vertical axis. For the sake of simplicity, we assume that the shock front is a plane and the MC is an uniform cuboid, and the distribution of the CRs inside the cloud is one-dimensional \citep{makino}. The CRs have been assumed to be confined in a region around the shock front with a width of W$_{sh}$, which is shown with a grey region in the figure. R$_{sh}$, r$_{MC}$ and R$_{esc}$ signifies the radius of the shock front, the distance of the surrounding MCs and the escaping radius from the cavity center respectively. The confined region moves outward with a velocity of  v$_{sh}$ and the overlapped region between the MC and the confinement region is given by d$_{MC}$. Part of the accelerated protons, marked with red circles escapes the confinement region through R$_{esc}$, and seeps inside the MC. These escaped protons further interact with with the cold protons inside the MC (marked with orange circles) and produce $\gamma$-rays.}
\end{figure*}

In this section, we calculate the hadronic contribution to the total $\gamma$-ray flux observed by LHAASO from LHAASO J1908+0621. The hadronic component comprises of the $\gamma$-ray produced from the interaction between escaped protons from SNR G40.5-0.5 and cold protons residing inside the associated MCs. We assume that the SNR and MCs are at a distance of 8 kpc from the Earth, similar to \cite{Li21}. For that distance, the number density of the associated MCs were assumed to be 45 cm$^{-3}$ \citep{Li21}. As evident by the radio observations, the SNR shows a shell like structure, outside of which the MCs are present. Due to this fact, we assumed that the supernova exploded at the center of the cavity of the shell which is surrounded by MCs, similar to \cite{fujita09}. After the explosion, the shock expands inside the cavity and finally hits the surrounding MCs, which is assumed to be $\sim$ 22 pc from the cavity center.

After the explosion, the supernova is in the free expansion phase, in which the ejecta from the explosion expands freely without any deceleration. After time t$_{Sedov}$, the supernova enters the adiabatic Sedov phase, in which mass of the swept-up ISM material by the shock wave increases and reaches densities which impedes the free expansion. Rayleigh-Taylor instabilities arise once the mass of the swept up ISM approaches that of the ejected material. In this phase, the cooling timescales is essentially much longer than the dynamical timescales, which makes this phase adiabatic in nature. This phase lasts until t$_{Rad}$, after which the supernova enters the radiative phase. When the shock expands through various phases of supernova evolution, its radius and velocity changes with time. The time dependence of shock velocity is given by \citep{ohira12, fujita09},

\begin{equation}
\label{eq1}
\begin{split}
v_{sh}(t) = 
\begin{cases}
v_i & (t < t_{Sedov})\\
v_i(t/t_{Sedov})^{-3/5} & (t_{Sedov} < t)\\
\end{cases}              
\end{split}
\end{equation}

where v$_{sh}$ is the velocity of the shock and v$_i$ denotes the initial velocity of the ejecta. We assume v$_i$ = 10$^9$ cm s$^{-1}$ \citep{fujita09}. We can obtain the time dependence of the shock radius by integrating equation \ref{eq1}.

\begin{equation}
\label{eq2}
\begin{split}
R_{sh}(t) \propto 
\begin{cases}
(t/t_{Sedov}) & (t < t_{Sedov})\\
(t/t_{Sedov})^{2/5} & (t_{Sedov} < t)\\
\end{cases}              
\end{split}
\end{equation}

For this work, we assume radius of the shock and time at the beginning of the Sedov phase, R$_{Sedov}$ and t$_{Sedov}$, to be 2.1 pc and 210 yr following \cite{ohira11, makino}.

The CR protons are accelerated through Diffusive Shock Acceleration (DSA) mechanism when the supernova is in the Sedov phase. The CRs are scattered back and forth across the shock front by magnetic turbulence during the acceleration as the shock front is expanding towards the surrounding MCs. Following \cite{ohira12, makino}, we assume CR protons need to cross an escape boundary outside the shock front to escape from the SNR. To that end, we assume a geometrical confinement condition l$_{esc}$ = $\kappa$R$_{sh}$ and adopt $\kappa$ = 0.04 \citep{makino}, where l$_{esc}$ is the distance of the escape boundary from the shock front. Using this definition and equation \ref{eq2}, we can write the escaping radius,

\begin{equation}
\label{eq3}
R_{esc}(t) = (1 + \kappa) R_{sh}(t),
\end{equation}

We assume that the accelerated CR protons need to cross this escaping radius to contribute to further astrophysical processes.  

After traversing through the cavity, the SNR shock eventually hits the surrounding MCs. The shock has to travel a distance of $\sim$ 22 pc (= r$_{MC}$, distance of the MCs from the cavity center) to collide with the MCs. Setting R$_{esc}$ = r$_{MC}$, from equation \ref{eq2} and \ref{eq3}, the time of the collision can be found to be t$_{coll} \sim$ 7.5 $\times$ 10$^3$ yrs. Using t$_{coll}$ in equation \ref{eq1}, the velocity of the shock at the point of collision can be calculated to be v$_{sh}$(t$_{coll}$) $\sim$ 1.2 $\times$ 10$^8$ cm/s. Following \cite{fujita09}, we assume the SNR is at the end of the Sedov phase at t = t$_{coll}$, so the particle acceleration stops at t $\sim$ t$_{coll}$. Hence the protons accelerated at t $\leq$ t$_{coll}$ ($\sim$ t$_{rad}$) will illuminate the MCs. However, in order to interact with the cold protons inside MCs, the accelerated CR protons have to escape from the SNR shock first. The protons which have higher energies will have more probability to escape the confinement region and take part in the hadronic interaction. Protons with lower energies will not be energetic enough to escape the confinement region. Since we are considering the interaction at a time when the outermost boundary of the confinement region (R$_{esc}$) collides with the MC surface (r$_{MC}$), i.e. t$_{coll}$, only the higher energy protons will take part in the hadronic interaction at t$_{coll}$, and the lower energy protons will still be confined around the SNR. Consequently, a dominant hadronic contribution primarily in the highest energy, while a suppression in the escaped proton population in the lower energies, is expected from this scenario. This condition not only puts a constraint on the lower energy limit of the escaped proton population from the SNR shock, but also on the spectral shape of the escaped protons. The CRs with higher energies escape the confinement region and start seeping into the MC when the escaping boundary (R$_{esc}$) contacts the surface of the MC (r$_{MC}$) \citep{makino}. The schematic diagram explaining the collision, as well as the escape of the accelerated proton population is shown in Figure \ref{fig1}.

To estimate the minimum energy needed to escape the SNR shock, we use a phenomenological model, where the escape energy is expected to be a decreasing function of the shock radius \citep{makino}. This approach is based on the assumption that SNRs are responsible for observed CRs below the knee \citep{gabici09, ohira12}. The maximum energy of CR protons E$_{max}$ is expected to increase up to the knee energy (= 10$^{15.5}$ eV) until the beginning of the Sedov phase, and then it decreases from that epoch. The escape energy can be given by a phenomenological power-law relation,

\begin{equation}
\label{eq4}
E_{esc} = E_{max} \left(\frac{R_{sh}}{R_{Sedov}}\right)^{-\alpha},   
\end{equation}

where $\alpha$ is a parameter describing the evolution of the maximum energy during the Sedov phase \citep{makino, ohira12}. In this paper, we assume that $\alpha$ = 2, which also dictates the suppression of escaped proton population at lower energies. Hence, assuming E$_{max}$ = 10$^{15.5}$ eV, R$_{sh}$ = R$_{esc}$ = r$_{MC}$ and R$_{Sedov}$ = 2.1 pc, we get the minimum energy needed by the CR protons to escape from the confinement region formed around the SNR shock front, when the escape boundary contacts the surrounding MC surface, i.e. E$_{esc}$ $\approx$ 30 TeV. We assume E$_{esc}$ = E$_{min}$ while calculating the total hadronic contribution from the escaped CR proton population. We can also calculate the spectral index of the escaped CR proton population to ascertain its spectral shape. Since the protons are accelerated by DSA mechanism, we assume that the CR proton spectrum at the shock front is represented by a power-law $\propto$ E$^{-s}$. Then the spectrum of the escaped protons, i.e. the protons having the energy greater than E$_{esc}$ is given by \citep{makino, ohira12},

\begin{equation}
\label{eq5}
    N_{esc}(E) \propto E^{-[s + (\beta/\alpha)]},
\end{equation}

where $\beta$ represents a thermal leakage model of CR injection and is given by $\beta$ = 3(3 - s)/2. For s = 2, we get $\beta$ = 1.5. Plugging in the value of s, $\alpha$ and $\beta$ in equation \ref{eq5}, we get the spectral index of the escaped CR protons to be $\approx$ 2.75. Note that the spectral shape and the minimum energy of the escaped protons are calculated when the escape boundary hits the surface of the surrounding MCs (t = t$_{coll}$).  

\begin{figure*}
\gridline{\fig{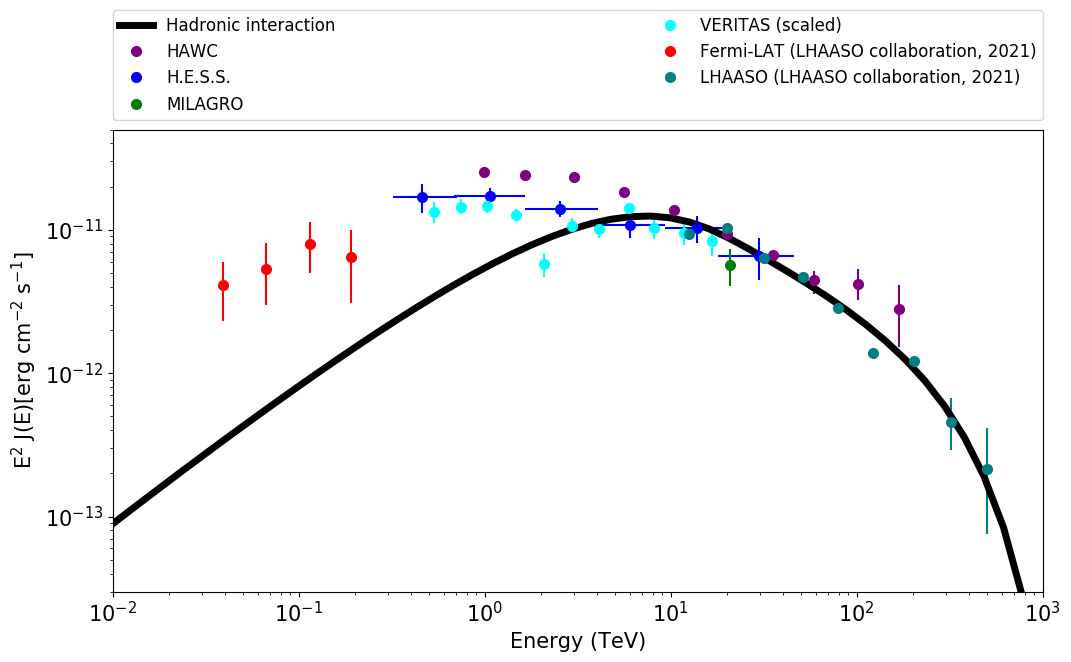}{0.5\textwidth}{(a)}
          \fig{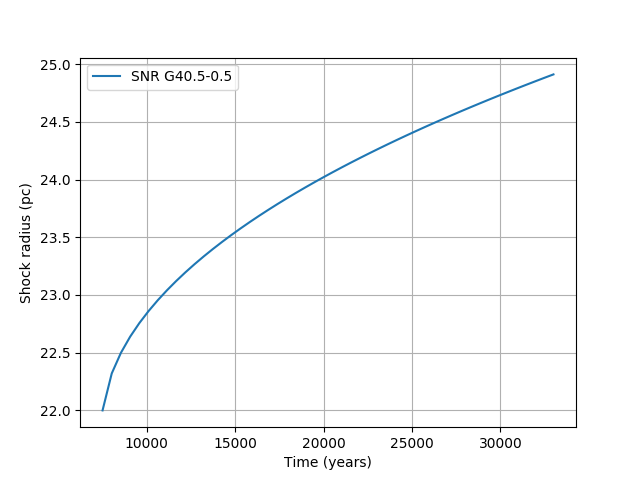}{0.5\textwidth}{(b)}
          }
          \caption{(a) Model $\gamma$-ray SED obtained from hadronic p-p interaction inside the MCs surrounding the SNR G40.5-0.5. Along with the calculated SED, datapoints obtained from \textit{Fermi}-LAT (red) \citep{Cao21}, VERITAS (cyan) \citep{aliu14}, H.E.S.S. (blue) \citep{aharo09}, MILAGRO (green) \citep{abdo07}, HAWC (purple) \citep{hawc19} and LHAASO (teal) \citep{Cao21} are also shown. The VERITAS data points have been scaled to match with that measured by H.E.S.S. observatory. (b) Time evolution of the shocked shell associated with the SNR G40.5-0.5, inside the surrounding MCs are shown.
\label{fig2}}
\end{figure*}

After the collision at t$_{coll}$, the shock enters the momentum conserving, radiative pressure driven ``snowplow'' phase of evolution at t $>$ t$_{coll}$. Similar to \cite{fujita09}, we can express the shock in the cloud as a shell centered on r = $|\overrightarrow{r}|$ = 0, where, $\overrightarrow{r}$ represents radially outward direction from the cavity center. Furthermore, from the momentum conservation, the radius of the shocked shell R$_{shell}$(t) inside the MCs can be written as \citep{fujita09}, 

\begin{equation}
\label{eq6}
\begin{split}
\frac{4\pi}{3}&\left[n_{MC} ( R_{shell}(t)^3 -  R_{sh}(t_{coll})^3) + n_{cav}R_{sh}(t_{coll})^3\right]\Dot{R}_{shell}(t)\\
& = \frac{4\pi}{3}n_{cav}R_{sh}(t_{coll})^3 v_{sh}(t_{coll}), 
\end{split}
\end{equation}

with R$_{shell}$ = r$_{MC}$ at t = t$_{coll}$. n$_{MC}$ = 45 cm$^{-3}$, is the number density of the associated MCs and n$_{cav}$ = 1 cm$^{-3}$ represents the number density inside the cavity, which we choose to be same as that of the interstellar medium (ISM). We solve the equation \ref{eq6} numerically at t $>$ t$_{coll}$. We found that the it takes the time t$_{stop}$ $\sim$ 3.3 $\times$ 10$^4$ kyr for the shell radius to reach the radius of the observed shocked shell radius of $\sim$ 25 pc \citep{yang06}, and the calculated shell velocity of the shocked shell inside the MCs was found to be $\sim$ 55 km s$^{-1}$, which is very close to the observed internal gas velocity of the clouds of 10 km s$^{-1}$ \citep{yang06}. t$_{stop}$, which essentially indicates the age of the SNR+MC system, agrees well with the current age of the SNR G40.5-0.5 (20 - 40 kyr), and also the shocked shell velocity agrees well with the observation. If the velocity of the shocked shell was equal to the internal gas velocity of the clouds, then no shell would have been observed. But since a shocked shell inside the surrounding MCs has been observed, it is expected for the velocity of the shell to be somewhat higher. This shows that the model is consistent with present day observations of SNR G40.5-0.5. The variation of the shocked shell with time has been given in the Figure \ref{fig2} (b).  

Now in this work, we have assumed that escaped CR protons that have entered clouds do not escape from the clouds before they lose energy through rapid radiative cooling \citep{makino, fujita09}. For that, the diffusion coefficient inside the MCs has to be very low compared to that observed in the ISM. Alternatively, it can be represented by the condition t$_{diff}$ $\geq$ t$_{stop}$, where t$_{diff}$ is the diffusion time of CRs inside a cloud, and it is given by t$_{diff}$ $\sim$ L$_{MC}^2$/6D(E). L$_{MC}$ is the size of the MCs and D(E) is the energy dependent diffusion coefficient \citep{fujita09}.

From the observed secondary-to-primary ratio of CR in the Galaxy, the energy dependent diffusion coefficient in the Galaxy has been found to be D(E) $\approx$ 10$^{28}\chi$(E/10 GeV)$^{\delta}$ cm$^2$s$^{-1}$, where $\delta$ can be between 0.3-0.6 and $\chi$ is a multiplicative factor \citep{desarkar21}. It has been estimated before that although the value of $\chi$ is 1 in the Galaxy, inside dense MCs, the value is $\chi$ $<$ 1  \citep{gabici09}. The small value of $\chi$ can be attributed to the reduction of diffusion coefficient by the plasma waves generated by stream of escaping CRs near the vicinity of the SNR \citep{wentzel, fujita09}. In order to fulfill the condition t$_{diff}$ $\geq$ t$_{stop}$, we found that the value of $\chi$ must follow the condition $\chi \leq$ 0.01 \citep{fujita09}. This suppression of diffusion coefficient can be easily realized inside a dense molecular cloud environment, as the value of the diffusion coefficient inside MCs is estimated to be of the order of 10$^{25}$ - 10$^{26}$ cm$^2$s$^{-1}$ \citep{gabici09}. If this is the case, then we can comfortably state that the injected CR protons inside the MCs lose their energy before escaping from the MCs. Moreover, since there is no effect of diffusion on the injected CR proton population, the spectral shape of the proton population does not change before they lose their energy radiatively. So we can assume the injected CR proton population attains steady-state before losing energy through hadronic p-p interaction. Thus, we calculate the total $\gamma$-ray produced from this proton population through hadronic p-p interaction, while keeping in mind that the $\gamma$-ray spectrum calculated at present age will be same as that calculated at t $\sim$ t$_{coll}$. 

We have used GAMERA \citep{hahn} to calculate the steady-state $\gamma$-ray spectra from the population of injected CR protons inside the MCs that surround the SNR G40.5-0.5. We have used a CR proton population having power law spectrum in the form of N$_p$ $\propto$ E$^{-\alpha_p}$, with spectral index of $\alpha_p$ $\approx$ 2.75, minimum energy of E$_{min}$ $\approx$ 30 TeV and maximum energy of E$_{max}$ $\approx$ 10$^{15.5}$ eV, i.e. the knee energy. We have considered the semi-analytical method developed by \cite{kafe} to perform the hadronic interaction calculation. The magnetic field inside the cloud was assumed to be B$_{MC}$ $\sim$ 60 $\mu$G \citep{fujita09} and the number density used was n$_{MC}$ = 45 cm$^{-3}$ \citep{Li21}. The total energy of the injected protons needed to fit the data observed by various observatories is W$_p$ $\sim$ 2.5 $\times$ 10$^{49}$ erg, which is consistent with the usual 1 - 10 $\%$ of the kinetic energy released in SNRs (E$_{SN}$ = 10$^{51}$ erg) \citep{aha04}. The calculated spectrum, alongwith the observed datapoints are given in Figure \ref{fig2} (a).

From the figure, it can be seen that the $\gamma$-ray data observed by LHAASO, HAWC, H.E.S.S. and VERITAS were partially explained by the hadronic model, due to the suppression of the parent proton population at sub-TeV energies. It is also evident from the figure that an additional emission component is required for explaining the GeV - TeV part of the SED observed by \textit{Fermi}-LAT, H.E.S.S. and VERITAS. Moreover, lower energy $\gamma$-ray datapoints (not shown in Figure \ref{fig2} (a)) obtained by \cite{Li21} could not be explained by the same hadronic model, further indicating the necessity of additional emission components. In the next section, we aim to explain the sub-TeV, as well as lower energy (0.1 - 10 GeV) $\gamma$-ray datapoints using leptonic contributions from both SNR G40.5-0.5 and the PWN associated with PSR 1907+0602.

\section{Leptonic modeling} \label{sec:lep}

\subsection{PWN J1907+0602}\label{subsec:PWN}

Along with the hadronic contribution discussed in section \ref{sec:had}, we also take into account the contribution from the leptonic emission of relativistic electron population from the PWN powered by the rotation powered GeV pulsar PSR J1907+0602. The offset between the centroid of this extended source and PSR J1907+0602 indicates that this is a relic PWN \citep{Li21}. We have considered a steady-state relativistic electron population from this PWN and calculated the total leptonic contribution from this source.

We have considered different leptonic cooling mechanisms, such as IC, synchrotron and bremsstrahlung \citep{blumenthal, ghisellini, baring}, and obtained the total $\gamma$-ray SED from the electron population associated with the PWN using GAMERA \citep{hahn}. The distance and the age of the PWN was set at 3.2 kpc and 19.5 kyr respectively \citep{abdo10}, same as that of PSR J1907+0602. The value of the magnetic field associated with a PWN, in general, is low ($\sim$ $\mu$G), due to the adiabatic expansion of the PWN with time \citep{martin12}. The magnetic field associated with PWN J1907+0602 was assumed to be B$_{PWN}$ $\approx$ 3 $\mu$G, in order to be consistent with previous works by \cite{Li21} and \cite{crestan}. The number density inside the PWN was assumed to be n$_{PWN}$ = 0.1 cm$^{-3}$. To calculate the IC contribution from the PWN, we have considered the ISRF model from \cite{popescu}. We have also considered the contribution from Cosmic Microwave Background (CMB), having the temperature T$_{CMB}$ = 2.7 K and energy density of U$_{CMB}$ = 0.25 eV cm$^{-3}$. The spectrum of the electron population was assumed to be a simple power law with exponential cutoff in the form of N$_e$ $\propto$ E$^{-\alpha^{PWN}_e}$ exp(-E/E$^{e, PWN}_{max}$). The spectral index of the spectrum was taken as $\alpha^{PWN}_e$ $\approx$ 1.5 and the maximum energy of the population was considered to be E$^{e, PWN}_{max}$ $\approx$ 10 TeV, which is constrained by the observed X-ray upper limits. The minimum energy of the electron population E$^{e, PWN}_{min}$ was given by the rest-mass energy. The energy budget of this relativistic electron population needed to satisfy the observed VHE data, was found to be W$^{PWN}_e$ $\sim$ 7.5 $\times$ 10$^{47}$ erg.

\begin{figure*}[ht!]
\plotone{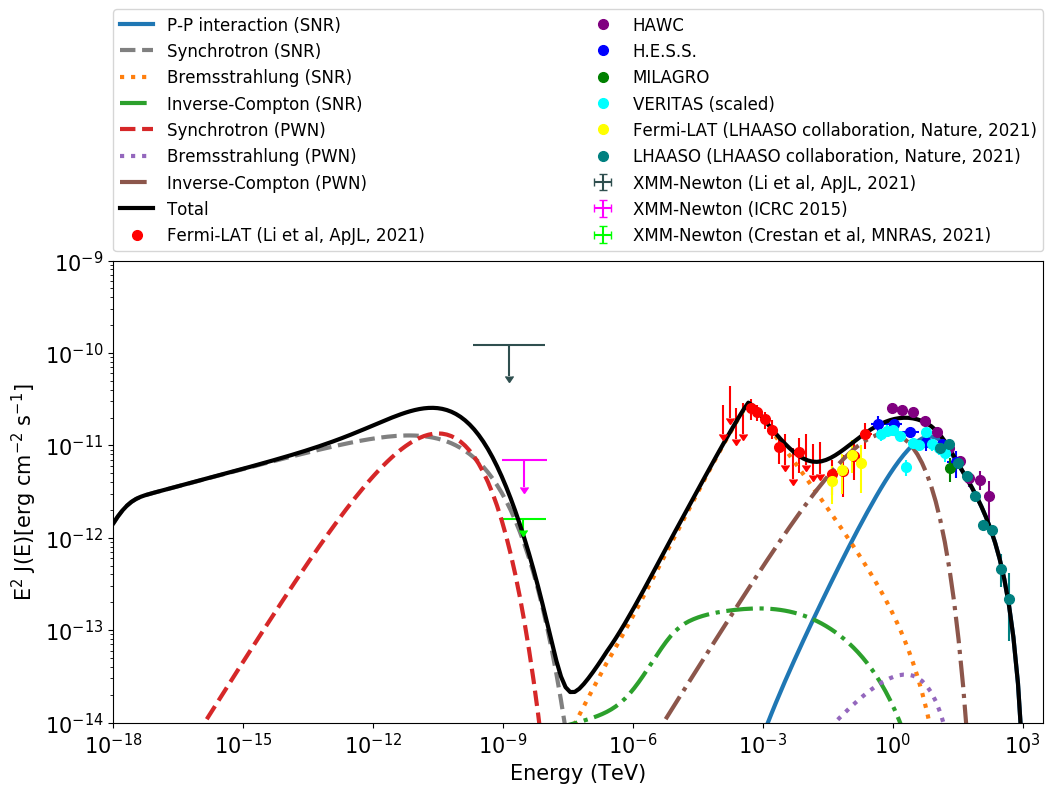}
\plotone{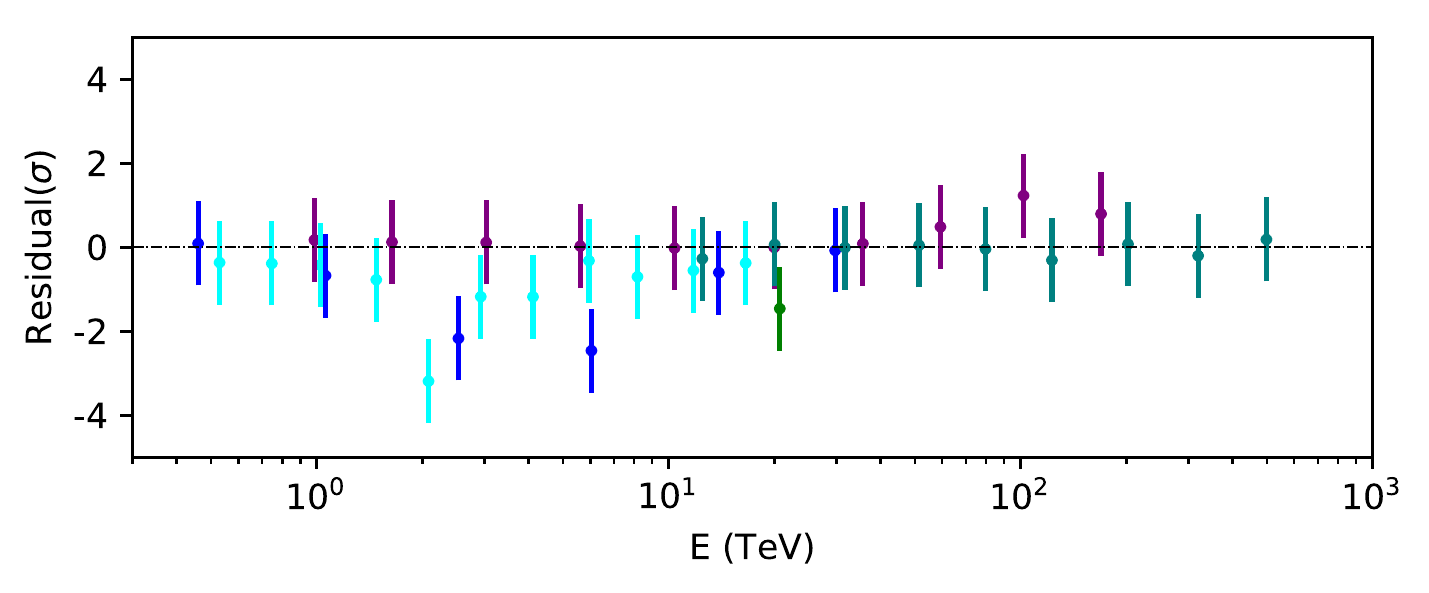}
\caption{\label{fig3} \textit{Upper Panel:} MWL SED of LHAASO J1908+0621. Datapoints obtained from different observations by \textit{Fermi}-LAT (red \citep{Li21}, yellow \citep{Cao21}), HAWC (purple) \citep{hawc19}, H.E.S.S. (blue) \citep{aharo09}, MILAGRO (green) \citep{abdo07}, VERITAS (cyan) \citep{aliu14} and LHAASO (teal) \citep{Cao21} are shown in the figure. The VERITAS data points have been scaled to match with that measured by H.E.S.S. observatory.. The XMM-Newton upper limit obtained from \cite{Li21} is shown in darkslategrey. XMM-Newton upper limits obtained from \cite{crestan} and \cite{pandel15} are shown in lime and magenta respectively. Solid blue line corresponds to the hadronic component from SNR G40.5-0.5. The synchrotron (grey dashed), bremsstrahlung (orange dotted), IC (light green dot-dashed) components from SNR G40.5-0.5 are shown. Also synchrotron (red dashed), bremsstrahlung (violet dotted), IC (brown dot-dashed) components from PWN J1907+0602 are shown. The total combination of all of these components are shown with a black solid line. \textit{Lower Panel:} The corresponding residual plot for the fit of the total model SED to the observed data from different observatories. The color scheme of the data points is same as that described in the \textit{Upper Panel}.}
\end{figure*}

\subsection{SNR G40.5-0.5}\label{subsec:SNR}

An extended object, labeled Fermi J1906+0626, illuminated in the GeV $\gamma$-ray range, was reported from the off-pulse phase resolved analysis done by \cite{Li21}. In their work, spectrum of this extended object, bright in the lower energy, was described by hadronic interaction between SNR+MC system, modified by the diffusion inside the clouds. However, in this work, we consider that the lower energy spectrum is due to the contribution from the electrons escaped from the shock front of SNR G40.5-0.5, a scenario which has not been explored in previous works. Escaped electron population from the confinement region around the shock, gets injected inside the surrounding MCs, and then interacts with the ambient medium of the same MCs. Emission is produced through synchrotron \citep{blumenthal, ghisellini} and IC \citep{blumenthal} cooling of the injected electron population. Since the number density of the MCs are much higher compared to that of ISM, bremsstrahlung emission \citep{baring} dominates the lower energy $\gamma$-ray SED. We have used GAMERA \citep{hahn} as before, to calculate the leptonic emissions.

Since the electrons go through the same evolution process as the protons before escaping from the SNR shock front, we assume that the spectral index of escaped CR electrons is same as that of the protons \citep{ohira12}, i.e. $\alpha^{SNR}_e$ $\approx$ 2.75. However, since the electrons, being leptons, lose energy radiatively very fast compared to protons, we have considered a simple power law with exponential cutoff as the spectrum of the runaway electron population in the form of N$_e$ $\propto$ E$^{-\alpha^{SNR}_e}$ exp(-E/E$^{e, SNR}_{max}$). The maximum energy associated with the runaway electron population spectrum is given by the relation \citep{yamazaki06, fujita09},

\begin{equation}
     \label{eq7}
     E^{e, SNR}_{max} = 14h^{-1/2}\left(\frac{v_{sh}}{10^8\:\rm cm/s}\right)\left(\frac{B}{10\:\rm \mu G}\right)^{-1/2} \rm TeV
\end{equation}

where, h($\sim$ 1) is determined by the shock angle and the gyro-factor, v$_{sh}$ is the velocity of the shock front and B is the downstream magnetic field. Since we calculate the maximum energy of the lepton population at the collision time, we considered v$_{sh}$(t$_{coll}$) as velocity in the above relation. We consider the magnetic field (B$_{MC}$) and number density (n$_{MC}$) inside the MCs same as that considered in section \ref{sec:had}, so we use B = B$_{MC}$ in the above relation. The minimum energy of the electron population was assumed to be E$^{e, SNR}_{min}$ = 500 MeV. For the IC contribution, we have adopted the interstellar radiation field (ISRF) modeled in \cite{popescu} at the position of SNR G40.5-0.5. The CMB contribution was also taken into account. The necessary energy budget of the runaway electron population to explain the lower energy $\gamma$-ray SED was found to be W$^{SNR}_e$ $\sim$ 1 $\times$ 10$^{49}$ erg. 

After considering synchrotron, IC and bremsstrahlung contributions from the runaway electron population from SNR G40.5-0.5, the lower energy (0.1 - 10 GeV) $\gamma$-ray SED obtained by \cite{Li21}, could be explained adequately. The dominant contribution in explaining the SED in 0.1 - 10 GeV range, came from the bremsstrahlung component, which is expected, as the morphology associated with this lower energy $\gamma$-ray emission was found to be spatially coincident with the molecular material enhancement observed inside the dense clumps surrounding SNR G40.5-0.5. The IC contribution was rather negligible in this case. The necessary model parameters used in this work have been summarized in Table \ref{tab1}.

\begin{deluxetable*}{ccccccccc}
\tablenum{1}
\tablecaption{\label{tab1} Parameters used in the model.}
\tablewidth{0pt}
\tablehead{
\colhead{Source} & \colhead{Component} & \colhead{Parameter} & \colhead{Value}}
\startdata
SNR G40.5-0.5 + MCs & Hadronic & Injection spectral index ($\alpha_p$) & 2.75\\
                                & & Minimum energy (E$_{min}$) & 30 TeV\\
                                & & Maximum energy (E$_{max}$) & 3.2 PeV\\
                                & & Energy budget (W$_p$) & 2.5 $\times$ 10$^{49}$ erg\\
                                & & Magnetic field (B$_{MC}$) & 60 $\mu$G\\
                                & & Number density (n$_{MC}$) & 45 cm$^{-3}$\\
                    & Leptonic & Injection spectral index ($\alpha_e^{SNR}$) & 2.75\\
                                & & Minimum energy (E$_{min}^{e, SNR}$) & 500 MeV\\
                                & & Maximum energy (E$_{max}^{e, SNR}$) & 6.9 TeV (Equation \ref{eq7})\\
                                & & Energy budget (W$_e^{SNR}$) & 1 $\times$ 10$^{49}$ erg\\
                                & & Magnetic field (B$_{MC}$) & 60 $\mu$G\\
                                & & Number density (n$_{MC}$) & 45 cm$^{-3}$\\
\hline
PWN J1907+0602 & Leptonic & Injection spectral index ($\alpha_e^{PWN}$) & 1.5 \\
                                & & Minimum energy (E$_{min}^{e, PWN}$) & 0.511 MeV\\
                                & & Maximum energy (E$_{max}^{e, PWN}$) & 10 TeV\\
                                & & Energy budget (W$_e^{PWN}$) & 7.5 $\times$ 10$^{47}$ erg\\
                                & & Magnetic field (B$_{PWN}$) & 3 $\mu$G\\
                                & & Number density (n$_{PWN}$) & 0.1 cm$^{-3}$\\
\enddata
\end{deluxetable*}

The MWL SED of the source LHAASO J1908+0621 is shown in the Figure \ref{fig3}, along with calculated SEDs from various leptonic and hadronic contributions from SNR G40.5-0.5 and PWN J1907+0602. From the figure, it can be seen that the total model flux satisfies the observed $\gamma$-ray SED data points, from lower energies to VHE-UHE regime.  Most notably, the UHE $\gamma$-ray spectrum, observed by LHAASO can be explained by the hadronic component from the SNR+MC system. \textit{Fermi-}LAT data points above 30 GeV, as obtained by \cite{Li21}, as well as \cite{Cao21}, were explained by the leptonic contribution from the PWN, which also conforms with the \textit{Fermi}-LAT morphology map obtained by \cite{Li21} (see Figure 2 of that paper).  Moreover, both PWN J1907+0602 and the SNR+MC system contribute in explaining the $\gamma$-ray SED observed by VERITAS and H.E.S.S. The bremsstrahlung emission from the escaped electron population associated with SNR G40.5-0.5, also satisfies the lower energy $\gamma$-ray SED. Very crucially, the combined synchrotron emission obtained from our model satisfies all of the upper limits obtained from various XMM-Newton data analysis \citep{Li21, crestan, pandel15}, further confirming the validity of our model. 

As can be seen from the upper panel of Figure \ref{fig3}, there remains a discrepancy between the data obtained by the Imaging Atmospheric Cherenkov Telescope (IACT) experiments and HAWC in the energy band of 1 to 10 TeV. Since HAWC has observed larger source extent \citep{abeysekara}, the data observed by both H.E.S.S. and VERITAS are inconsistent with that observed by HAWC \citep{crestan}. Consequently, in this particular work, we have tried to fit the data in this important band of 1-10 TeV, by favoring more the HAWC data than the IACT data. The corresponding residual, i.e. (data-model)/error, plot is given in the lower panel of Figure \ref{fig3}. From the figure, it can be clearly seen that the total model SED is more consistent with the HAWC data, as compared to the data observed by H.E.S.S. and VERITAS in the 1-10 TeV range.

\section{Neutrino flux }\label{sec:neu}

\begin{figure}
\plotone{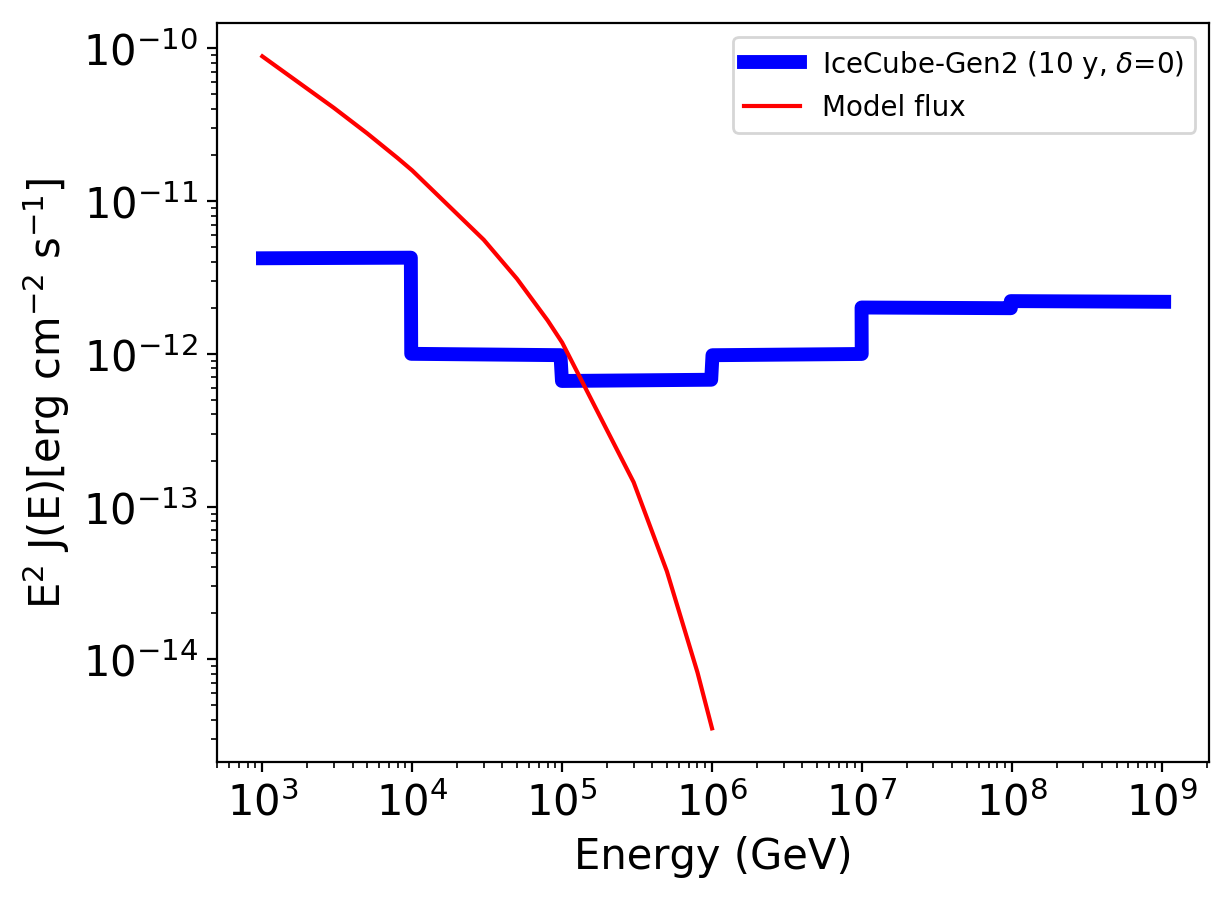}
\caption{The estimated total muonic neutrino flux reaching the Earth from SNR G40.5-0.5. The red continuous line represents total muonic neutrino flux produced due to the interactions of the escaped CR protons from SNR G40.5-0.5 with the cold protons in the associated molecular clouds. The blue solid dashed line indicates the sensitivity of IceCube-Gen2 to detect the neutrino flux from a point source at the celestial equator with an average significance of 5$\sigma$ after 10 years of observations. }
\label{fig4}
\end{figure}

Neutrinos are also produced in hadronic p-p interactions, along with $\gamma$-rays. Consequently, if there are $\gamma$-ray sources which are powered by hadronic interactions, neutrino emission from the same source region is also expected. MGRO J1908+06, the MILAGRO counterpart of LHAASO J1908+0621, may be a neutrino source due to its extended nature and hard TeV $\gamma$-ray spectrum \citep{gonza, halzen}. ICECUBE neutrino telescope searched for point-like source emission in the vicinity of this source. The astrophysical muon neutrino flux observed from this source region, has been found to have the second best p-value, being a Galactic source \citep{icecube19}. However, the emission is still consistent with the background. Although not quite significant yet, the presence of a neutrino hotspot associated with the source indicates hadronic emission in the highest energy range. The hadronic p-p interaction considered in the SNR+MC system to explain the UHE $\gamma$-rays observed by LHAASO, also produces neutrinos in the source region. In this section, we calculate the total muonic neutrino flux produced from the interactions between the escaped CR parent proton population from the SNR G40.5-0.5 and the cold protons residing inside the surrounding MCs.

To calculate the flux of the muonic neutrinos $\nu_{\mu}$ + $\Tilde{\nu}_{\mu}$, we use the semi-analytical formulation developed in \cite{kelner}. Following \cite{kelner}, we have included the muonic neutrinos produced from direct decay of charged pions ($\pi$ $\rightarrow$ $\mu$ $\nu_{\mu}$) labeled as $\nu^{(1)}_{\mu}$ and from the decay of muons ($\mu \rightarrow$ e $\nu_{\mu}$ $\nu_e$) labeled as $\nu^{(2)}_{\mu}$. E$_p$ and E$_\nu$ denote the energies of the proton population and produced neutrinos respectively. The total neutrino production rate from inelastic hadronic p-p interaction can be calculated using the equation \citep{kelner},

\begin{equation}
\label{eq8}
    \Phi_\nu (E_\nu) = \frac{cn_{MC}}{4\pi d^2}\int\sigma_{inel}(E_\nu/x) J_p(E_\nu/x) F_\nu(x, E_\nu/x) \frac{dx}{x} 
\end{equation}

where the variable x = E$_\nu$/E$_p$,  c is the velocity of light, n$_{MC}$ is the density of the molecular clouds, d(=8 kpc) is the distance of the SNR+MC system, $\sigma_{inel}$(E$_p$) is the inelastic cross-section of p-p interaction, which is given by,

\begin{equation}
\label{eq9}
\sigma_{inel}(E_p) = 34.3 + 1.88L + 0.25L^2\:\rm mb
\end{equation}
where L = ln(E$_p$/1 TeV). J$_p$(E$_p$) signifies the spectrum of the parent proton population and it is given by J$_p$(E$_p$) = A(E$_p$/1 TeV)$^{-\alpha_p}$. The normalization A (in the unit of erg$^{-1}$) can be calculated by performing the integration W$_p$ = A $\int^{E_{max}}_{E_{min}}$ E$_p$ (E$_p$/1 TeV)$^{-\alpha_p}$ dE$_p$, where the integration parameters are the same as that discussed in section \ref{sec:had}. The condition that the maximum energy of the parent proton population can reach up to E$_{max}$ $\approx$ 10$^{15.5}$ eV, is taken into account while considering the proton spectrum. F$_\nu$ represents the function that explains the spectra of $\nu^{(1)}_{\mu}$ and $\nu^{(2)}_{\mu}$, which get produced from the decays of charged pions and muons respectively \citep{kelner}. Note that while for $\nu^{(2)}_{\mu}$, the lower and upper integration limits for equation \ref{eq8} are 0 and 1 respectively, for $\nu^{(1)}_{\mu}$, the upper limit is 0.427. This is because the spectrum of F$_{\nu^{(1)}_{\mu}}$ sharply cuts off at x = 0.427 \citep{kelner}. By integrating equation \ref{eq8} with appropriate limits, we get the total muonic neutrino flux $\nu_{\mu}$ + $\Tilde{\nu}_{\mu}$, obtained from both channels of decays, and it is given by $\Phi_{\nu_{\mu} + \Tilde{\nu}_{\mu}}$ = $\Phi_{\nu^{(1)}_{\mu} + \Tilde{\nu}^{(1)}_{\mu}}$ + $\Phi_{\nu^{(2)}_{\mu} + \Tilde{\nu}^{(2)}_{\mu}}$.
Our estimated muon neutrino flux is shown in Figure \ref{fig4} along with the ICECUBE-Gen2 sensitivity limit \citep{icecubegen2}.

From Figure \ref{fig4}, it can be seen that the model neutrino flux exceeds the sensitivity limit of ICECUBE-Gen2. This implies that if the hadronic component from the SNR+MC system contributes to the total observed emission from the direction of LHAASO J1908+0621 in TeV - PeV range, then the corresponding neutrino flux will be detectable by ICECUBE below PeV energies. This is an important differentiator between the leptonic and hadronic scenarios in the UHE range, as no neutrinos would get produced if the UHE $\gamma$-ray emission from the LHAASO source is due to the IC cooling of one-zone or two-zone leptonic population \citep{hawc21}. Future observations by ICECUBE will help to confirm the exact nature of LHAASO J1908+0621.

\section{Discussion \label{sec:DC}}

In earlier literature \citep{abdo10, Li21}, it has been posited that the multi-TeV, VHE-UHE $\gamma$-ray data points are most likely represented entirely by the leptonic emission from a population of relativistic electrons associated with the PWN of PSR J1907+0602. \cite{Li21} have explained the VHE-UHE $\gamma$-ray data points with leptonic component from the PWN, since the spectrum measured by Cherenkov instruments resemble the spectral signature associated with IC emission from GeV/TeV PWNe. This approach was also considered in \cite{crestan}, as well as in \cite{hawc21}. However, \cite{crestan} disfavored a one-zone leptonic scenario to explain the VHE-UHE $\gamma$-ray spectra by the spatial morphology of the multi-TeV emission. The multi-TeV emission region associated with the PWN extends far from the pulsar position, but emission itself does not show any signature of spectral softening with the distance from the pulsar, as is expected from the cooling of electrons \citep{crestan}. Moreover, due to the Klein-Nishina suppression of the IC cross-section at higher energies, the maximum energy of the electron population will attain a large value, in order to fit the observed VHE-UHE $\gamma$-ray data entirely by a one-zone electron population from PWN J1907+0602. Furthermore, if the multi-TeV, VHE-UHE $\gamma$-ray datapoints were explained with a one-zone leptonic model from PWN J1907+0602, then the corresponding synchrotron flux would be incompatible with the X-ray upper limits obtained by \citep{crestan, pandel15} in the keV energy range. These issues were echoed in \cite{hawc21}, in which the authors also favoured a two population scenario to explain the VHE-UHE $\gamma$-ray emission. Note that \cite{crestan} also explored a one-zone hadronic model to explain the VHE-UHE $\gamma$-rays and found that a very hard photon index is needed in their model, which was not seen in other TeV sources associated with SNRs. Hence, they concluded that a fully hadronic model is also disfavoured. \cite{hawc21} also stated that a one-zone hadronic model to explain the UHE $\gamma$-rays observed by HAWC is not favoured due to lack of sufficient energy to power the hadronic emission and fit the observed data. It is clear that in order to explain the VHE-UHE $\gamma$-ray emission, a two population model is required. Any one-zone leptonic, as well as hadronic scenario is not sufficient for explaining the MWL SED of LHAASO J1908+0621 consistently.
 
In this paper, we explore a lepto-hadronic scenario of CR interaction to produce VHE-UHE $\gamma$-rays observed from the direction of LHAASO J1908+0621, with a particular focus on proper hadronic modeling required to explain both UHE $\gamma$-rays observed by LHAASO, as well as the neutrino hotspot coincident with the source position, detected by ICECUBE. We have considered that the emission in 10 GeV - 10 TeV energy range has originated due to the leptonic emission from PWN J1907+0602, whereas above 10 TeV, the emission has a hadronic origin. We use a physically viable and detailed model of CR interaction inside a SNR+MC system \citep{fujita09, ohira12, makino} to partially explain the observed UHE $\gamma$-ray data points. The choice of the free parameter $\alpha$ constrains the minimum energy and the spectrum of the escaped CR proton population, which consistently reproduce the $\gamma$-ray SED in the multi-TeV energy range. Moreover, the model matches the present day observation of the state of the shocked shell inside the MCs. In addition, we have also included emission due to leptonic cooling from PWN J1907+0602. Finally, we have shown that by considering these two scenarios, the $\gamma$-ray data points extending from 10 GeV to 1 PeV, can be explained well.

\cite{Li21} had shown previously that in the 0.1 - 10 GeV range, an extended source, Fermi J1906+0626, is present overlapping the source region of MGRO J1908+06. The authors had described SED from this source by a soft spectrum, which is similar to the ones observed in evolved SNRs. Moreover, they reported that from the \textit{Fermi}-LAT analysis of this extended source, a significant peak was found coinciding with an enhancement of molecular cloud material, thus justifying the tentative hadronic origin of this low energy component. They further fitted the SED with a steep power law proton spectrum, which is modified by diffusion. The same model was considered in \cite{hawc21}, where lower energy data points were fitted by a hadronic component from the SNR+MC system. However, in this work, we explain the $\gamma$-ray SED in 0.1 - 10 GeV range, with leptonic contribution from the SNR G40.5-0.5 and its surrounding MC system. We considered a power law spectrum with an exponential cutoff to explain the relativistic electron population associated with the SNR+MC system. We considered that, like protons, electrons can also escape from confinement region around the shock front of the SNR, and get injected into the MCs. After considering various leptonic cooling mechanisms inside the MCs, we found that the leptonic component from SNR G40.5-0.5 is adequate to explain the lower energy $\gamma$-ray emission. Since bremsstrahlung emission dominates the IC cooling inside a dense molecular medium, it was primarily used to explain the lower energy $\gamma$-ray SED. Moreover, this emission scenario was also corroborated by the spatial morphology observed by \textit{Fermi}-LAT \citep{Li21}. The combined lepto-hadronic scenario explored in our model not only satisfactorily explains the observed $\gamma$-ray SED from low to ultra high energy, the synchrotron emission obtained from our model is also consistent with the observed X-ray upper limits.

Since there is a possibility of an ICECUBE neutrino hotspot present in the source region, we also calculated total muonic neutrino flux from the hadronic interaction considered in this paper. If the UHE emission is due to leptonic emission, then no neutrino would be seen from the source region. Although the hotspot is not significant yet, as observed by previous generation ICECUBE, the ICECUBE-Gen2 has a better sensitivity for detecting neutrino from a Galactic source. From our calculation, we found that the total neutrino flux exceeds the sensitivity limit of ICECUBE-Gen2, which implies that if the emission from LHAASO J1908+0621 is partially hadronic in origin, then ICECUBE-Gen2 will be able to detect neutrino from the source region. Future observation by ICECUBE will be crucial to divide the two emission contributions from SNR+MC system and the PWN currently considered.

Although the model explored in this paper satisfies the observed $\gamma$-ray data points, as well as the X-ray upper limits, a lot of issues are still needed to be clarified by future experiments in the MWL bands. Since the source region is very complex, with the PWN and the SNR+MC system juxtaposed within the 68$\%$ containment region of many observatories, further morphological observations are very crucial to better constrain the model. More detailed morphological observations in UHE $\gamma$-ray regime, which we expect that Cherenkov Telescope Array (CTA) will provide in the near future, will be important in discerning the source localizations, as well as their contributions. Moreover, long term X-ray and radio observations are very crucial to constrain the modeling of this source. X-ray and radio observations will not only constrain the magnetic field, it will also affect the minimum energy, the injection spectral index and the energy budget of the parent lepton population of both PWN and SNR+MC system. Furthermore, astrophysical neutrino detection at the source region will in turn confirm the contribution of the hadronic component from the SNR+MC system at the UHE regime.

\section{Conclusion \label{sec:CON}}

In conclusion, in this paper, we have studied the underlying emission mechanism of LHAASO J1908+0621 suggested by the observed data, and subsequently explored a simple, analytical, phenomenological model that is compatible with the MWL data points hitherto observed. In our model, the leptonic component from PWN J1907+0602 is dominant in 10 GeV to 10 TeV range, whereas the hadronic component is used to explain the observed UHE SED above 10 TeV to 1 PeV energy range. The leptonic contribution from the SNR+MC system explains the lower energy part (0.1 - 10 GeV) of the $\gamma$-ray SED. Our model also satisfies the observed X-ray upper limits. However, as discussed earlier, more detailed observations about the $\gamma$-ray emitters in energy ranges from 0.5 TeV to 1 PeV, will reveal more insight of the complex source region in short future. Additionally, the crucial observations in X-ray and radio bands will play a huge role in unveiling the radiation mechanism of this source region through detailed MWL analyses. Future observations by CTA observatory, as well as neutrino observation by ICECUBE-Gen2 at the source position will be important to untangle the exact nature of this enigmatic source in both low and high energies.

\begin{acknowledgments}
The authors thank the anonymous reviewer for constructive criticism and useful suggestions regarding the manuscript. A.D.S. thanks Maheswar Swar for help regarding Figure \ref{fig1}.
\end{acknowledgments}

\vspace{5mm}

\software{GAMERA (\url{https://github.com/libgamera/GAMERA})
          }

\bibliography{sample631}{}
\bibliographystyle{aasjournal}

\end{document}